\documentclass[amsmath,amsfonts,amssymb,twocolumn,nofootinbib]{revtex4-1}
\usepackage{graphicx}
\usepackage{verbatim}
\usepackage{hyperref}
\usepackage{xcolor}
\usepackage{subfigure}

\begin{document}

\title{\large{Limit setting using spacings in the presence of unknown backgrounds}}
\author{\normalsize{Lolian Shtembari}}
\email{lolian@mpp.mpg.de}
\author{\normalsize{Allen Caldwell}}
\affiliation{Max Planck Institute for Physics, Munich, DE 80805}

\begin{abstract}

\noindent Finding upper limits on the rate of events from a proposed process in the presence of unknown backgrounds is an often encountered problem in the search for rare processes.
Methods based on unusually large ``gaps'', or spacings, in the event distribution allow to set limits on the rate of the proposed signal distribution.
In this paper, we present two novel spacings-based methods: the ``Sum of sorted spacings'' and the ``Product of complementary spacings'' as  tests and compare these to existing tests on synthetic data as well as on a published data set.

\end{abstract}

\maketitle

\section*{Introduction}

\noindent Many experiments tasked with the discovery of theorised rare processes might find themselves in a situation where the collected data is insufficient to claim a positive detection. 
In such cases, the data is used to set an upper limit on the number of events resulting from the rare process under consideration, which in turn can be used to set an upper limit on physical quantities of the proposed model.
An example would be the determinations of upper limits on the cross-section of Weakly Interacting Massive Particles (WIMPs) recoiling off atoms in a detector, such as for the CRESST \citep{CRESST:1999ynq} or the CDMS \citep{Abrams_2002} experiments. The experiments in question are often contaminated by a poorly understood background, in which case the signal strength limit must be set from properties of the observed event distribution without any background subtraction.  Spacing statistics are one method to go beyond pure event counting in setting signal strength limits.

Since the expected shape of the event distribution produced by the targeted process is known, it is possible to estimate the number of events it accounts for, up to a desired confidence level, leveraging the difference between the observed event distribution and the expected one.
Such an analysis is carried out using goodness-of-fit tests allowing for the assumption that the observed number of events collected in the analysis window is just a realization of a random variable following a Poisson distribution with unknown rate $\mu$.
For a selected goodness-of-fit test, the goal is to determine the event rate $\mu$ coinciding with the desired confidence level.

In the following we give a brief review on how to use goodness-of-fit tests for an arbitrary event distribution assuming the targeted process and how to account for the random number of observed events in the definition of the p-value.

We then discuss how to use spacings-based tests to provide upper limits: we begin with a quick review of the Maximum Gap and Optimum Interval methods \citep{Yellin:2002} and then introduce two new tests based respectively on the sorted list of spacings and on the product of spacings.

\section*{Setting upper limits with test statistics}

\noindent 
Several non-parametric goodness-of-fit tests exist to test a univariate distribution.
Targeting vastly different univariate distributions is made possible by the probability integral transformation \citep{pearson_1902, pearson_1933}, which basically reduces the goodness-of-fit to a simple uniformity test.

\subsection*{Probability integral transformation}

\noindent Suppose we have $n$ samples $\{y_i\}$, and want to quantitatively test the hypothesis of those samples being random variates of a known distribution $f(y)$, independent and identically distributed (i.i.d.) according to $f(y)$. 
Considering only continuous distributions $f(y)$ with cumulative $F(y)$, it is possible to transform samples onto the unit interval $[0,1]$ via $x_i = F(y_i)$. 
This reduces the task at hand to test transformed samples $\{x_i\}$ being distributed according to the standard uniform distribution $\mathcal{U}(0,1)$.

\subsection*{Poisson distribution and p-value}

\noindent Given a dataset consisting of $n$ uniformly distributed samples $\{x_i\}$, a test statistic $T$ and a function of the data $t = g(\{x_i\})$, the p-value is directly calculable from $F_T(t | n)$, where $F_T$ is the cumulative distribution function of the test $T$ for exactly $n$ events.  Here, the p-value treats the number of events in the analysis window as a fixed parameter. 
If we consider that the observed number of events is a random variable with an associated Poisson distribution, then it is possible to correct the definition of the p-value by averaging over all possible numbers of events.
Considering a Poisson distribution with rate $\mu$, and an observed test statistic value 
$t_{obs} = g(\{x_i\})$, the Poisson-averaged p-value is calculated as:

\begin{equation}
    1 - p = F_{T,Pois}(t_{obs} | \mu) = \sum_{n = 1}^{\infty} F_T(t_{obs} | n) \cdot \frac{\mu^n e^{-\mu}}{n!}
\end{equation}

\noindent where the sum starts at $n=1$ since the test statistic often is not defined for $n=0$. 
In case of no observed events, $n=0$, no improvement upon the simple Poisson statistic is possible, which becomes the extension of this approach in the limit of empty datasets.

\subsection*{Setting upper limits}

\noindent Given a test $T$ and its distribution $F_{T,Pois}$, we showed above how to calculate the p-value of a given dataset $\{x_i\}$ comprised of $n$ events, for a selected value of the event rate $\mu$. 
Instead of performing a goodness of fit test, assessing how well the event distribution fits that of a uniform distribution for a given $\mu$, we could determine which is the rate $\mu$ that yields a desired p-value, determining the event rate representative of the uniformly distributed subset of events up to a desired confidence level ($CL$).

As an example, for a given dataset $\{x_i\}$ the upper limit on the event rate, $\mu_U$, at a confidence level $CL$ is such that:

\begin{equation}
    1 - p = F_{T,Pois}(t_{obs} | \mu_U) = CL
\end{equation}

\section*{Spacing statistics}

\noindent Given a uniformly distributed and ordered set of $n$ events $\{x_i\}$ (where $x_{i} < x_{j}\, \forall i<j$), in the interval $[0,1]$, we can define the $n+1$ spacings $s$ as $s_{j,1} = x_{j} - x_{j-1}$, with $x_{0} = 0$ and $x_{n+1}=1$.
These spacings are the gaps between consecutive events, but in general we could also consider higher order spacings: a spacing of order $k$ would be $s_{j,k} = x_{j} - x_{j-k}$.

Based on these spacings it is possible to construct test statistics capable of setting much more competitive upper limits on the event rate than the simple counting (Poisson) test, since they not only consider the total number of data contained in the analysis window, but also their distribution, taking advantage of regions of relatively low event density in order to estimate the underlying uniform component of the event distribution.

\subsection*{Maximum Gap method}

\noindent This Maximum Gap test has been proposed in the physics literature \citep{Yellin:2002} (and earlier in the statistics literature \citep{Fisher:1929} ).  It considers the largest spacing present in order to determine the upper limit on the event rate.
The test statistic is defined as:

\begin{equation}
    T_{MG}(\{x_i\}) = s_{max, 1} = \max_{i}(s_{i,1})
\end{equation}

\noindent The distribution of $s_{max, 1}$ for a given number of events is known \citep{Fisher:1929, Yellin:2002}.
 Under the assumption of a uniform distribution of events, a large proposed event rate, $\mu$, will lead to a small probability to observe large values of $s_{max, 1}$. If the observed value of $s_{max, 1}$ is indeed large relative to our expectations, this can be used to exclude values of $\mu$ at a specified confidence level.

This definition of the test is particularly helpful since it is little affected by possible clustering of events in the unit interval, whose distribution deviates from the standard uniform one.
This case arises when an experiment is afflicted by an unknown background and the probability integral transformation is performed with respect to the distribution of the signal under study: after the transformation the events deriving from a possible signal source will be uniformly distributed while events coming from unknown background sources will not follow a uniform distribution, producing regions in the unit interval with and over-density of events and regions with large spacings.

The Poisson averaged cumulative distribution, $F_{MG,Pois}$, can be easily computed analytically and has been given by Yellin \citep{Yellin:2002}:

\begin{equation}
    F_{MG,Pois}(x | \mu) = \sum_{t=0}^{m} \frac{(tx - \mu)^t e^{-tx}}{t!} \left( 
 1 + \frac{t}{\mu - tx}\right)
\end{equation}

\noindent where $m$ is the greatest integer $\leq \mu / x$.
The $CL$ upper limit on the event rate, $\mu_{MG}$, is such that:

\begin{equation}
    F_{MG,Pois}(s_{max, 1} | \mu_{MG}) = CL
\end{equation}

\subsection*{Optimum Interval method}

\noindent In addition to the Maximum gap method, the Optimum Interval method has been proposed by Yellin \citep{Yellin:2002} which instead of looking at the largest spacing, instead considers sums of ordered spacings: i.e., higher order spacings. 

The Maximum Gap method compares the size of the largest first order spacing against the expectation of it containing no events for a given event rate $\mu$.
Similarly, given higher order spacings, for example $k=2$, we might find the largest second order spacing $s_{max, 2}$ and compare its size to the expectation of it containing only one event given a proposed event rate $\mu$. 
Such an investigation can be performed for any order of spacings allowed by the data ($k \leq n$, where $n$ is the number of events) and would result in $n$ different Poisson-averaged p-values, one for each order of spacing, for a given event rate $\mu$:

\begin{align}
    & s_{max, k} = \max_i(s_{i,k}) \\
    & 1 - p_k = F_{O,k,Pois}(s_{max,k} | \mu)
\end{align}

\noindent where $F_{O,k}$ is the cumulative distribution of the $s_{max, k}$ for a given number of events.
The analytic formula of $F_{O,k}$ ($k>1$) for $n$ events is not known, but Yellin calculated numerical approximations using a large Monte Carlo campaign as well as an approximate asymptotic distribution for large values of $n$ \citep{Yellin:2007}.

In order to exclude the proposed event rate $\mu$, one might look at the smallest available p-value:

\begin{equation}
    p_{min} = \min_k(p_k)
\end{equation}

\noindent Since $p_{min}$ does not have a uniform distribution, one needs to know its cumulative distribution for a given event rate $\mu$, $F_{O, min}$, and the final p-value is:

\begin{equation}
    p_{fin} = F_{O, min}(p_{min} | \mu)
\end{equation}

\noindent The analytic formula of $F_{O, min}$ is not known, and a numerical approximation is derived using Monte Carlo simulations.
Inverting the formula, the upper limit $\mu_{OI}$ on the event rate up to a given $CL$ is such that:

\begin{equation}
    F_{O, min}(p_{min} | \mu_{OI}) = 1 - CL
\end{equation}

\subsection*{Sum of Sorted Spacings method}

\noindent The tests discussed so far were developed in order to be sensitive to the presence of any abnormally large gaps between the events.
For relatively few events under analysis, there might be only one such abnormally large gap, in which case the Maximum Gap method might already provide the most stringent upper limit.  If more than one such abnormally large gap is present, and
if these gaps happen to be located near one another, then they can be integrated in one higher order spacing and the Optimum Interval method provides more competitive limits.

However, if the abnormally large gaps are interspaced by many small gaps, then the Optimum Interval method would lose sensitivity and not offer significant improvements compared to the Maximum Gap method: in these cases, the sum of ordered spacings approach saturates and is dominated by individual low order spacings.

Ideally, we would like to combine the abnormally large gaps individually, without the gaps placed between them. 
This approach would increase the sensitivity of the test and potentially allow setting more competitive upper limits.  We now describe such an approach.

Given a set of $n$ events $\{x_i\}$ in the unit interval $[0,1]$, one considers the set of first order spacings $\{s_{i,1}\}$ and then proceeds to sort these from largest to smallest, constructing the set of $n+1$ elements $\{g_i\}$ (where $g_1 > g_2 > .. g_n > g_{n+1}$).
Given such a set of sorted spacings, $\{g_i\}$, it is possible to consider higher order spacings summing over its elements.
The $k$-th ordered sum of sorted spacings, $G_k$, is just the sum of the $k$ largest first order spacings:

\begin{equation}
    G_k = \sum_{i=1}^{k} g_i
\end{equation}

\noindent In total there are up to $n$ non trivial $G_k$ given $n$ events, since the sum over all $n+1$ first order spacings is constrained to be equal to 1.

\noindent It is now possible to evaluate the p-value of each order of sum of sorted spacings for a given event rate $\mu$:

\begin{equation}
    1 - p_k = F_{S,k,Pois}(G_k | \mu)
\end{equation}

\noindent where $F_{S,k}$ is the cumulative distribution of the $G_k$ for a given number of events.
The analytic formula of $F_{S,k}$ for $n$ events is known: it was first derived by Mauldon \citep{mauldon_1951}.  We independently re-derived it using an alternative approach reported in Appendix A.

As with the Optimum Interval method, it is possible to use the smallest p-value, $p_{min}$, as a test statistic in order to exclude a value of $\mu$ that is too large:

\begin{equation*}
    p_{min} = \min_k(p_k)
\end{equation*}

\noindent Since $p_{min}$ is not a valid p-value any more, one needs to know its cumulative distribution for a given event rate $\mu$, $F_{S, min}$, and the final p-value is calculated in this case as:

\begin{equation}
    p_{fin} = F_{S, min}(p_{min} | \mu)
\end{equation}

\noindent The analytic formula of $F_{S, min}$ is not known, and a numerical approximation is found from Monte Carlo simulations.
Inverting the formula, one needs to find the $CL$ upper limit $\mu_{S}$ such that:

\begin{equation}
    F_{S, min}(p_{min} | \mu_{S}) =  1 - CL
\end{equation}

Although the analytic formula of $F_{S,k}$ for $n$ is available, it is not well-behaved, since it relies on the iterative difference of extremely large numbers (as $n$ increases), making it susceptible to catastrophic numerical cancellation when used on a computer.
In order to avoid these problems, we computed the values of the function with high numerical precision on a suitable grid in order to construct a reliable monotonic cubic-spline interpolation \citep{doi:10.1137/0717021} that can be used with default 64-bit floating-point arithmetic.
Currently, these interpolations have been tabulated up to $n=700$ and allow the estimation of event rates up to $\mu \lesssim 580$, correspondingly limiting the number of events it is possible to analyse.
The speed-up tables and the code used to compute the test-statistic and the limits is available in the SpacingStatistics.jl \citep{SpacingStatistics.jl} package for Julia.

\subsection*{Product of Complementary Spacings method}

\noindent Finally, we propose another test statistic that combines spacings between events regardless of their relative location.
The stepping stone of our proposal is a test first proposed by Moran \citep{Moran:1951} which consists in the product of all the spacings between consecutive events:

\begin{equation}
    T(\{x_i\}) = - \sum_{i=1}^{n+1} \log(s_{i,1})
\end{equation}

\noindent where $s_{i,1}$ are the first order spacings and $n$ is the number of events in the unit interval.
This test was proposed as a goodness-of-fit test sensitive to clusters of data against the null-hypothesis of a uniform distribution: the presence of small spacings will drive the whole product of spacings towards more extreme values.

In order to make this test sensitive to the presence of large spacings, we consider the complements of each first order spacing and take their product:

\begin{equation}
    C(\{x_i\}) = - \sum_{i=1}^{n+1} \log(1 - s_{i,1})
\label{eq:product_of_complementary_spacings}
\end{equation}

\noindent The distribution of this quantity for a fixed number of events $n$, $F_C(C | n)$, is not known analytically, but we derived a numerical approximation based on Monte Carlo simulations and tabulated them for $n \leq 10^4$.
Additionally, since the definition of the test is of the form $\sum g(s_{i,1})$, it is in general possible to derive its asymptotic distribution, as described by Darling \citep{10.1214/aoms/1177729030} and LeCam \citep{LeCam:1958}, whose methods we used to find the limiting distribution of $C$.
The asymptotic distribution of $C$ as $n \rightarrow \infty$ is:

\begin{equation}
    f_C(C | n \rightarrow \infty) = \mathcal{N}(n \cdot \mu_{\infty},\, n \cdot \sigma_{\infty})
\end{equation}

\noindent where the formula of $\mu_{\infty}$ and $\sigma_{\infty}$ are reported in Appendix B.
Given this result we can use it to estimate the test statistic for any large value of $n$ and effectively extend the applicability of this test and its limit calculation to large numbers of events.

The Poisson-averaged p-value of this test for a given event rate $\mu$ is simply:

\begin{equation}
    F_{C, Pois}(C | \mu) = \sum_{n=1}^{\infty} F_C(C | n) \cdot \frac{\mu^n e^{-\mu}}{n!}
\end{equation}

\noindent Inverting this formula, one finds the $CL$ upper limit $\mu_{C}$ such that:

\begin{equation}
    F_{C, Pois}(C | \mu_{C}) = CL
\end{equation}

The tabulated distributions as well as the code used to compute the test-statistic and the limits is available in the SpacingStatistics.jl \citep{SpacingStatistics.jl} package for Julia.

\section*{Performance comparison}

\noindent The performance of our proposed methods was extensively studied using simulated examples. 
In our tests we introduced backgrounds of varying  shapes and strengths. 

In the following we compare our methods against the standard Poisson test and the Optimum Interval method, which is considered the state of the art for setting limits in experiments affected by an unknown background.

\subsection*{Background-free experiment}

\begin{figure}[h]
\centering
\includegraphics[width=0.45\textwidth]{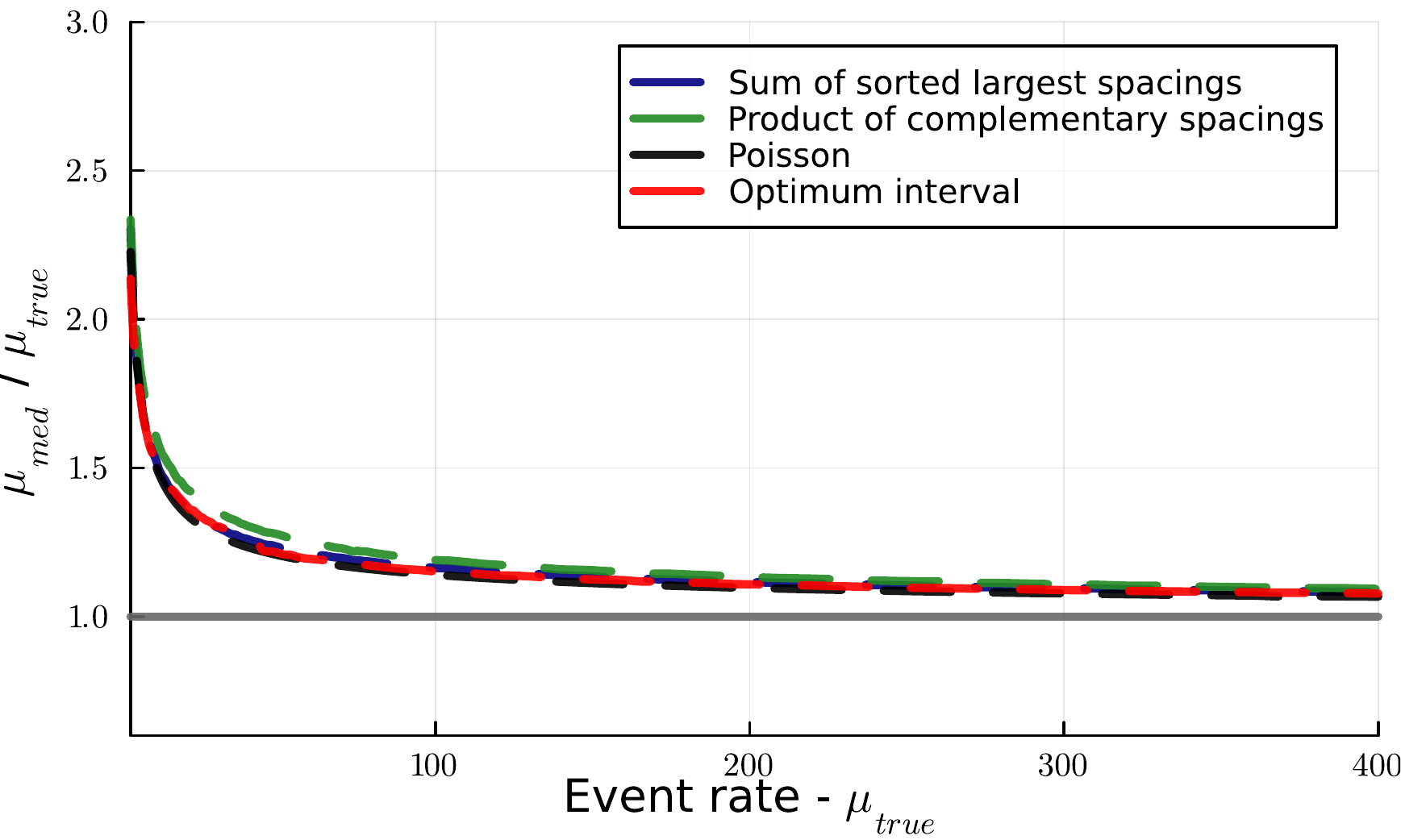}
\caption{Median $CL=0.90$  upper limit normalized to the event rate used in the simulations; data generated according to a uniform distribution (background-free).}
\label{fig:lim_no_bkg}
\end{figure}

\begin{figure}[h]
\centering
\includegraphics[width=0.45\textwidth]{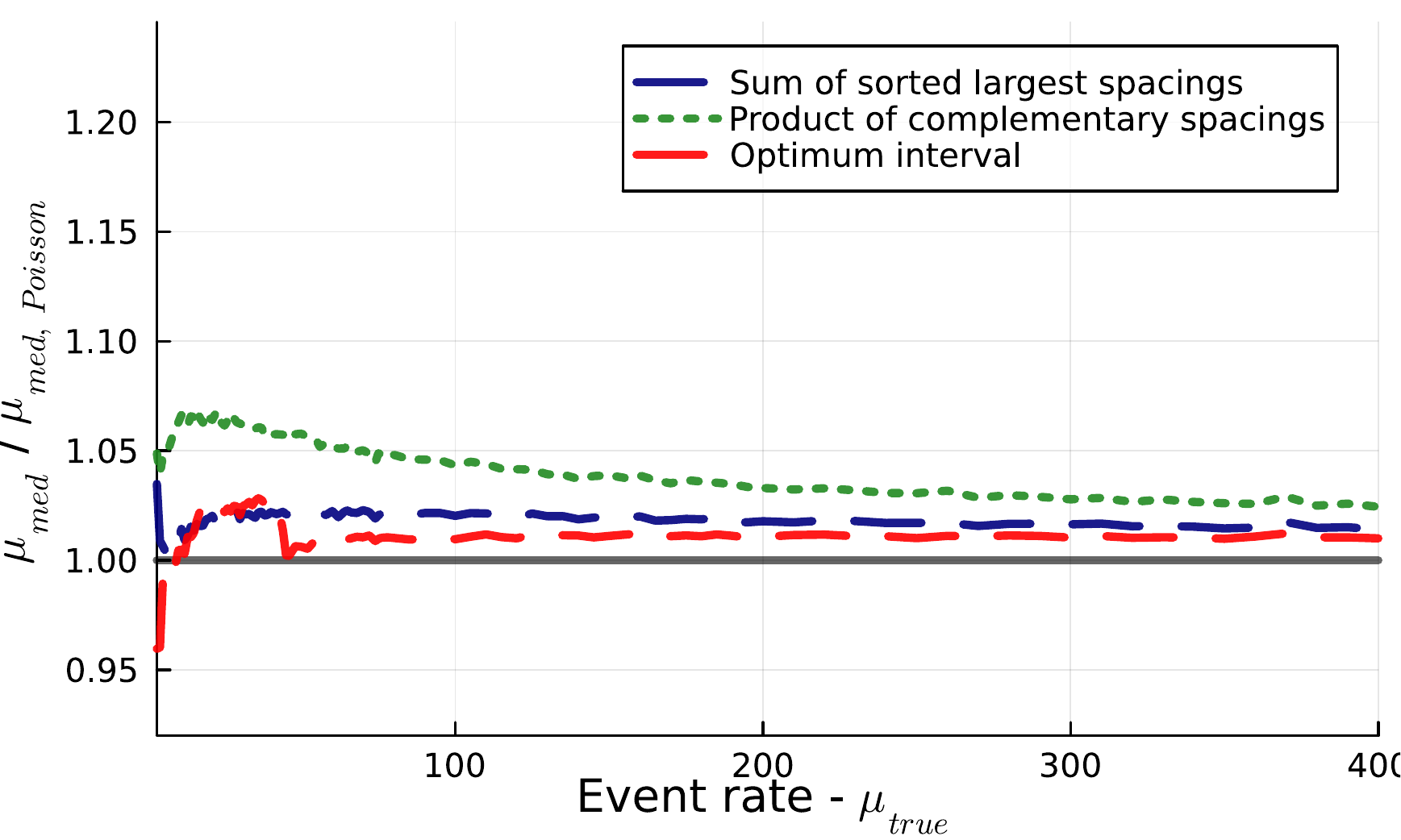}
\caption{Median $CL=0.90$ upper limit normalized to the Poisson-test's result; data generated according to a purely Uniform distribution (background-free).}
\label{fig:lim_no_bkg_rel_Yell}
\end{figure}

\noindent We start considering the case in which no added background contaminates the experiment, in order to estimate the baseline of the different methods. 
For simplicity we chose a uniform distribution for the generation of the events, which coincides with the null-hypothesis of all tests, and we vary the event rate used in the data generation.
Fig~\ref{fig:lim_no_bkg} shows the median of the $CL=0.90$ upper limits on the event rate set using different methods.
In order to better discern differences between the efficiency of each method, we can look at the results normalized to the Poisson limit, as shown in Fig.~\ref{fig:lim_no_bkg_rel_Yell}.
We notice that in this baseline scenario the Poisson test is the best of the bunch, as expected.
Nevertheless, the results of the Poisson test do not drastically outperform any other.


\subsection*{Exponential background-only experiment}

\begin{figure}[h]
\centering
\includegraphics[width=0.45\textwidth]{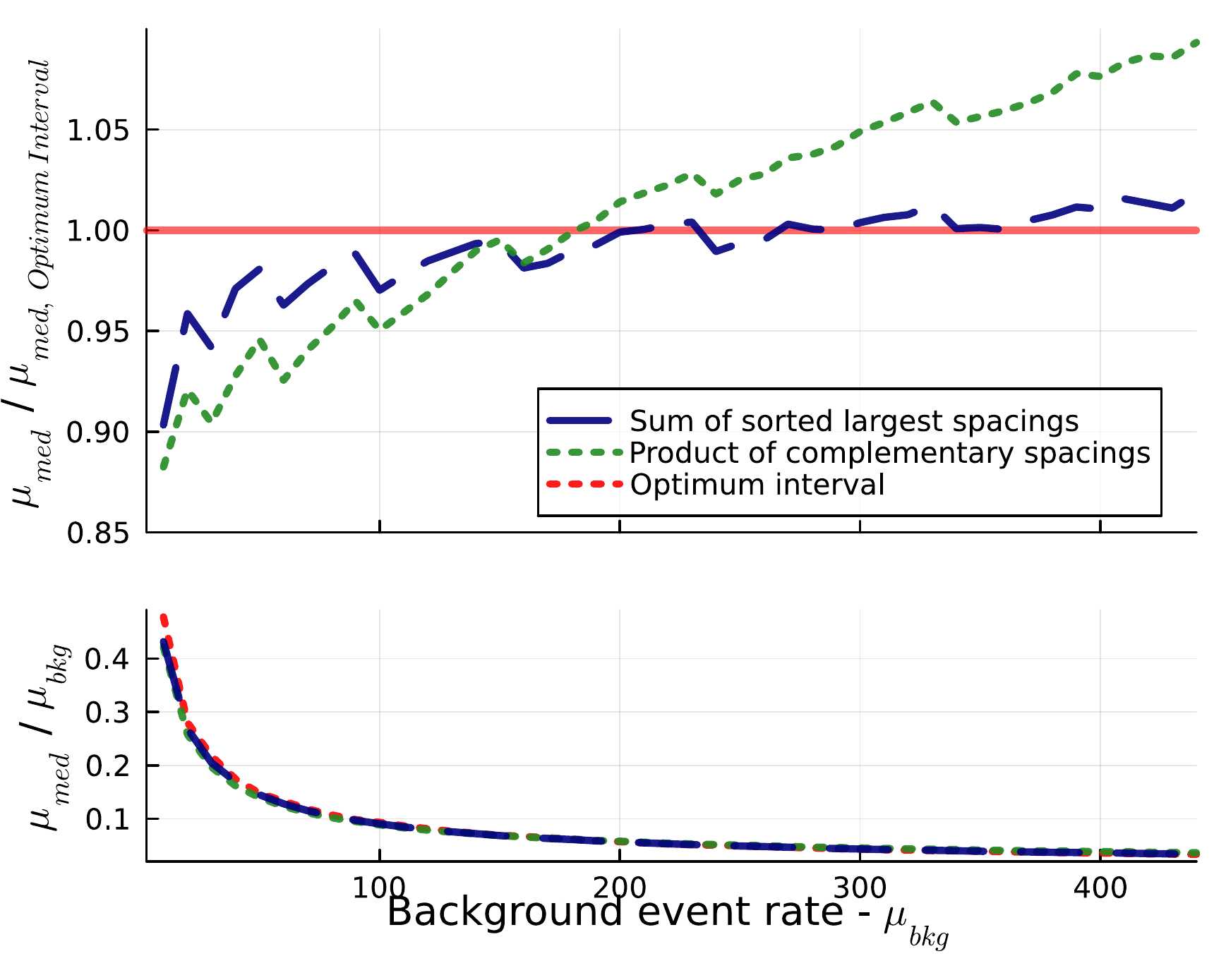}
\caption{Median $CL=0.90$ upper limit normalized to the Optimum Interval result; data generated according to an Exponential distribution of rate 0.1 (only-background).}
\label{fig:lim_no_sig_rel_Yell}
\end{figure}

\noindent Next we investigate the case in which a background is present in our simulations and the signal strength is negligible in comparison: this mimics a rare process search in which the signal might be absent.
In our experiments we produce data directly in the cumulative space (hence the signal distribution is always assumed to be flat, i.e. the null-hypothesis) and we start by considering an exponential background with rate 0.1 truncated on the unit interval $[0,1]$.
In dark matter search experiments it is often the case that the distribution of events, after transforming to the cumulative space, is peaked at one end of the analysis window with rapidly decaying tails.
Fig.~\ref{fig:lim_no_sig_rel_Yell} reports the median $CL=0.90$ upper limits of the measured event rate normalized to the injected background event rate (bottom) as well as the ratio of the results of our methods against the Optimum Interval test (top). 
We have omitted the Poisson limit as it would simply scale with the total number of events and is not competitive.
Analysing the results, we notice that when dealing with relatively peaked event distributions all methods perform similarly, just like the background-free case.
All methods are able to filter out most of the background contribution and reconstruct small overall event rates.
For small injected background rates ($\leq 100$), the non-local methods (Sum of sorted spacings and product of complementary spacings) are able to set up to $5-10$\% more stringent limits.
As the background rate increases ($\geq 200$), the performance of the Sum of sorted spacings' test matches the Optimum Interval's one, while the Product of spacings's results are up to $5-10$\% worse.

\subsection*{Mixing background and signal}

\begin{figure*}
\centering
\includegraphics[width=0.9\textwidth]{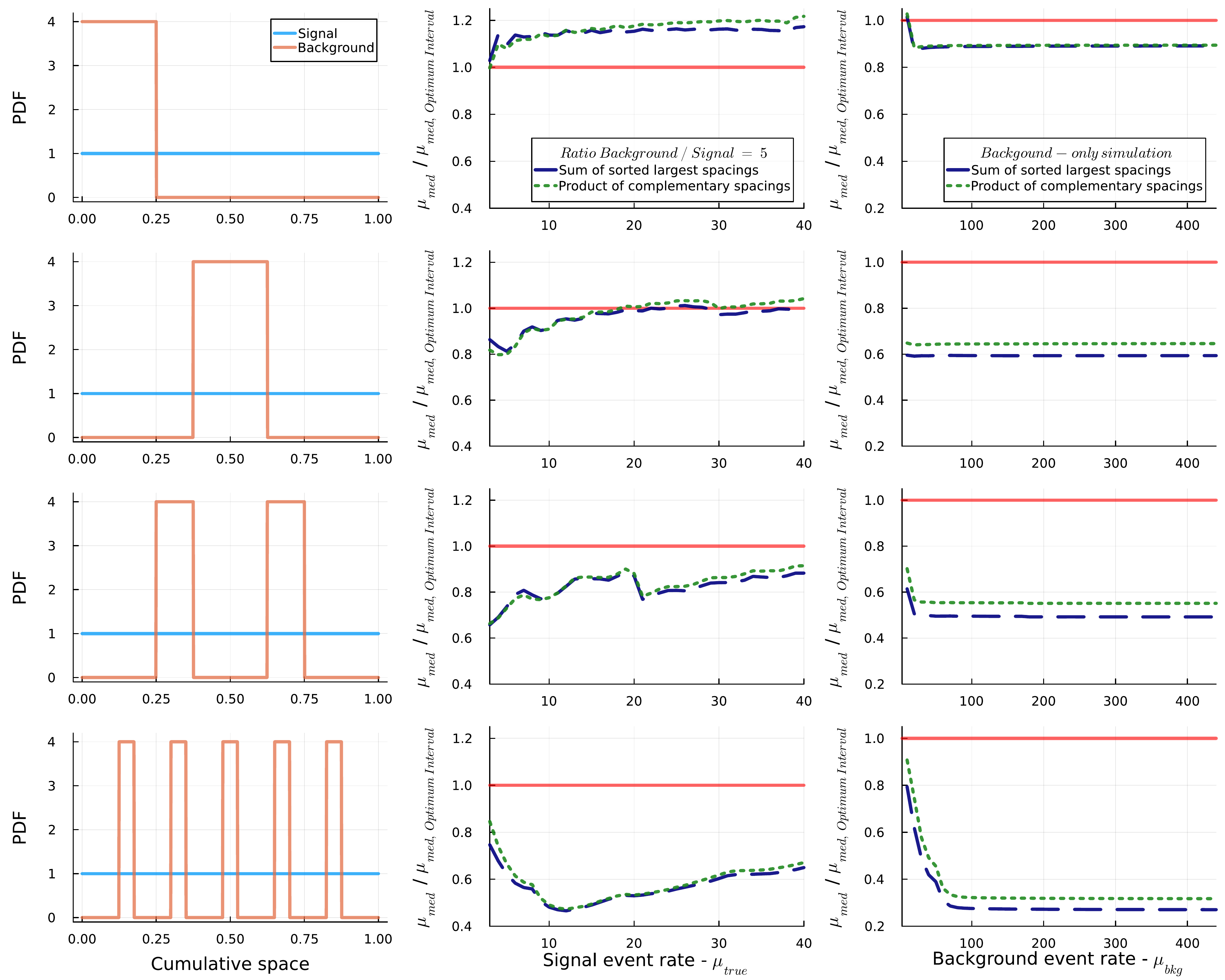}
\caption{Comparison of limit-setting methods depending on event distribution in the cumulative space: (left column) background and signal event distribution; (middle column) median $CL=0.90$ upper limit normalized to the Optimum Interval's result for simulated event distributions with a background and signal mixing of $\mu_{bkg} / \mu_{sig} = 5$; (right column) median 90\% CL upper limit normalized to the Optimum Interval's result for purely background-like event distributions ($\mu_{bkg} / \mu_{sig} = \infty$).}
\label{fig:bkg_5}
\end{figure*}

\noindent Now we investigate the case in which a detectable signal distribution is contaminated by  background.
In our simulations we will consider a background to signal ratio of 5, i.e the rate of background events is 5 times larger than the rate of signal events.
Since we operate directly in the cumulative space, we always assume a uniform signal distribution.
The support of the background event distribution spans a quarter of the analysis window (meaning that the background events occupy a quarter of the unit interval) and we consider distributions of varying shape: specifically, the background distribution is a mixture of one or more uniform distributions whose total width sums up to $0.25$. 

Fig.~\ref{fig:bkg_5} shows different choices of background distributions in the left column, the resulting median $CL=0.90$ limits for different methods, normalized to the Optimum Interval's result, in the central column.

If the background distribution if fully concentrated in one region, localised at either end of the analysis window, as shown in the first row of Fig.~\ref{fig:bkg_5}, this creates an uninterrupted low density region of the resulting event distribution.
This is the best case scenario for the Optimum Interval method, as previously discussed. 
This expectation is reflected in the results, where the Optimum Interval method's results are up to 20\% better than the other methods.

As we move the background distribution in the middle of the analysis window, or even split it up in two or more peaks, then we notice how the our proposed tests are more sensitive, being able to set more competitive limits.
For a bimodal background distribution it is possible to set limits 20\% lower than the Optimum Interval method on average, while the gain rises up to 40\% for a pentamodal background distribution.
The performance of our proposed tests (middle column of Fig.~\ref{fig:bkg_5}) is due to their sensitivity to large regions of low event density, regardless of their number or location.
With just two background peaks it is possible to being able to produce limits up to 20\% or even 40\% better than the Optimum interval method as the low event density regions are further splits.

Considering the case of a very faint or absent signal, we analyse the resulting limit if events were distributed only according to the background distribution.
The results of these simulations are shown on the right column of Fig.~\ref{fig:bkg_5}, where we notice that, regardless of the shape of the background distributions, the 90\% CL median results of the Sum-of-Sorted-Spacings and Product-of-Complementary-Spacings are always smaller than the Optimum Interval counterpart, with limit gains increasing up to a factor of 3 as the number of event-free regions increases.

The case of multimodal background distributions, especially when it presents well defined and relatively narrow peaks, is interesting since it is similar to experimental scenarios in which the event rate of a three or multi-body decay is sought after: the expected spectrum of such a decay is relatively flat and could be contaminated by peaking background distributions which are representative of processes with a Standard-Model counterpart. 
Our proposed methods would be well-suited to tackle these problems since they are able to filter out the contributions coming from these ``peaks'' and estimate the underlying ``flat'' event rate, without introducing additional parameters in the analysis (biasing the result) to modify, limit or segment the Region-of-Interest in order to exclude peaking backgrounds.

\subsection*{Comparison to a Likelihood-Ratio Test}

\begin{figure}[h]
\centering
\includegraphics[width=0.45\textwidth]{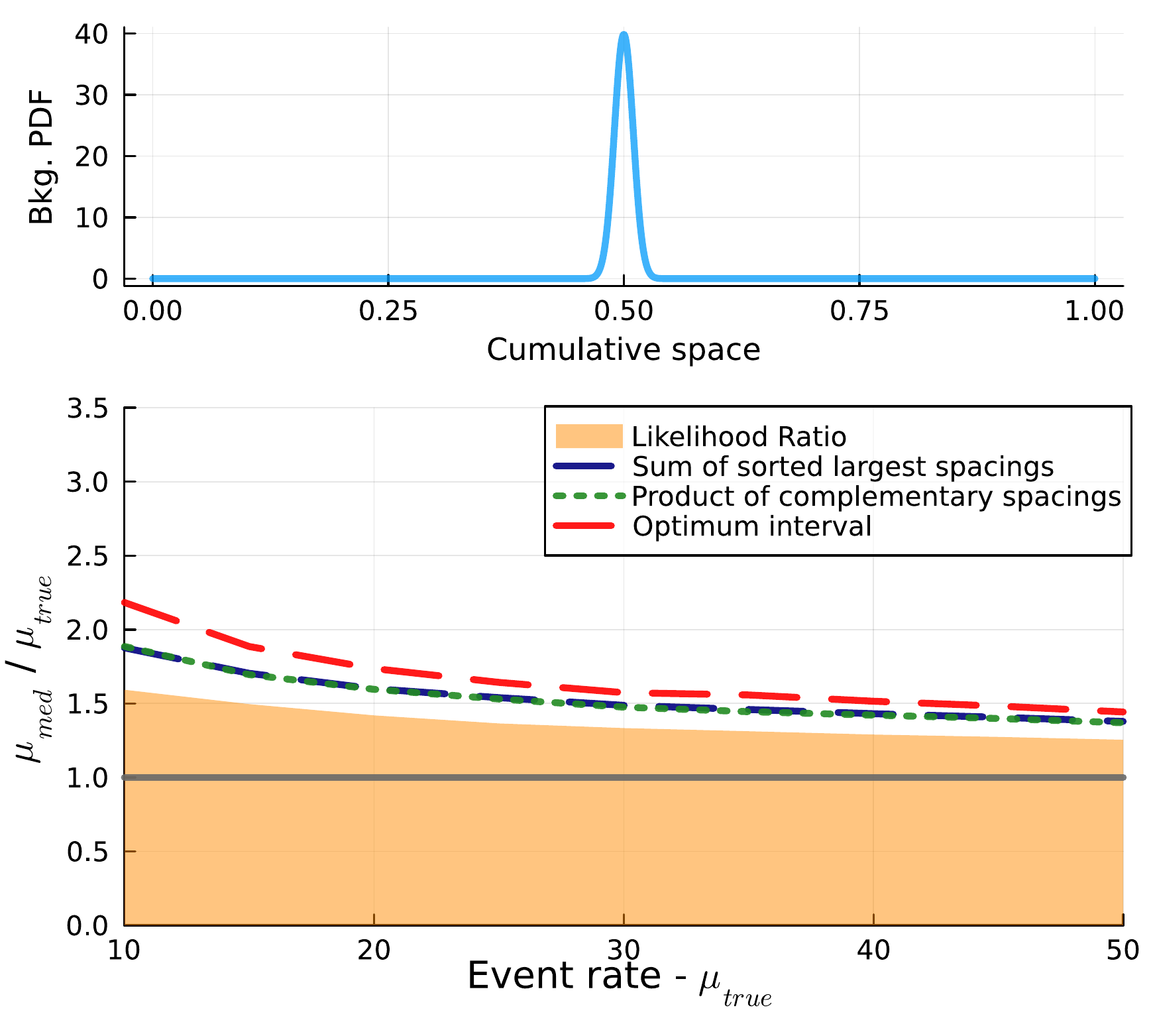}
\caption{(top) Background distribution in the cumulative space: Gaussian distribution with $\mu=0.5$ and $\sigma=0.01$; (bottom) Median $CL=0.90$ upper limit normalized to the signal event rate.}
\label{fig:likelihood_ratio}
\end{figure}

\noindent As a further example we compare the efficiency of the non-parametric tests discussed so far against a Likelihood-Ratio (LR) test in the case of peaking backgrounds.
We consider a Gaussian background with an associated event rate ten times stronger than the signal's event rate.
The original distribution family of the background was fed into the LR method while the position, width and the background to signal ratio were left as free parameters.
Fig.~\ref{fig:likelihood_ratio} shows the background distribution used and the median $CL=0.90$ upper limits obtained at different signal event rates.
Inspecting the results we notice that the most stringent limits are set by the LR approach, which is hardly surprising since partial information of the background was folded into the analysis.
The limits set by the non-parametric spacings based tests, although more conservative, are still close enough to the LR ones, with our methods providing results no more than 20\% larger than the LR for low event rates and up to 10\% larger limits for high even rates.
Furthermore, no assumption on the background shape is needed in the non-parametric tests.

\section*{CRESST Data Example}

\begin{figure}[h]
\centering
\includegraphics[width=0.45\textwidth]{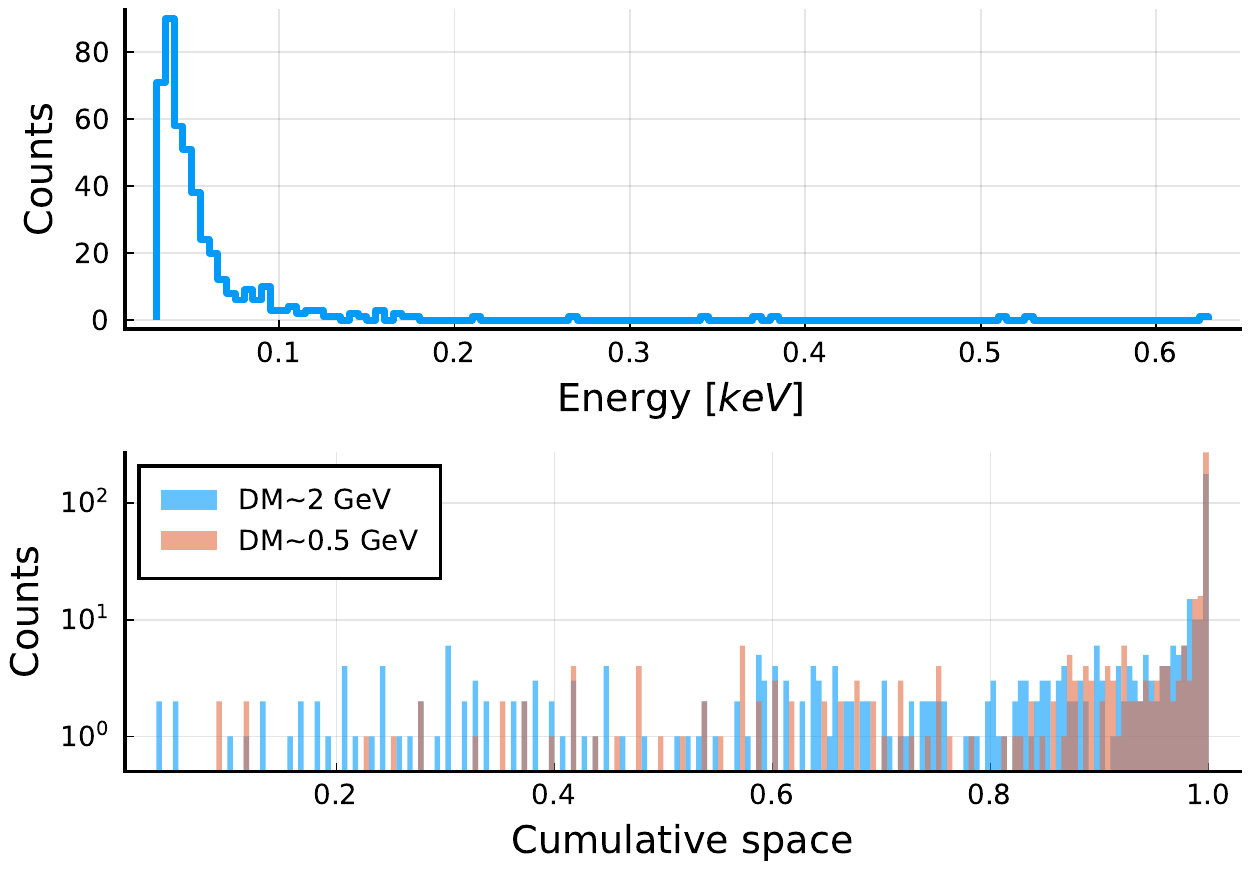}
\caption{(top) Histogram of CRESST data \citep{PhysRevD.100.102002} consisting of energy deposition from an interaction of a particle in the $\mathrm{CaWO_4}$ crystal; (bottom) Histogram of data transformed using the signal distribution for two proposed WIMP masses, $0.5$ and $2 \; \mathrm{Gev/c^2}$}
\label{fig:CRESST_data_keV}
\end{figure}

\noindent We now replicate the analysis of the CRESST Collaboration~\cite{CRESST:1999ynq} to determine upper limits on the WIMP cross-section.
We use the most recent public dataset \citep{PhysRevD.100.102002} released from the CRESST collaboration that is accompanied with information regarding the energy resolution and efficiencies of their setup for $\mathrm{CaWO_4}$ targets.
The data we analyse are shown in Fig.~\ref{fig:CRESST_data_keV} (top).

\begin{figure}[h]
\centering
\includegraphics[width=0.45\textwidth]{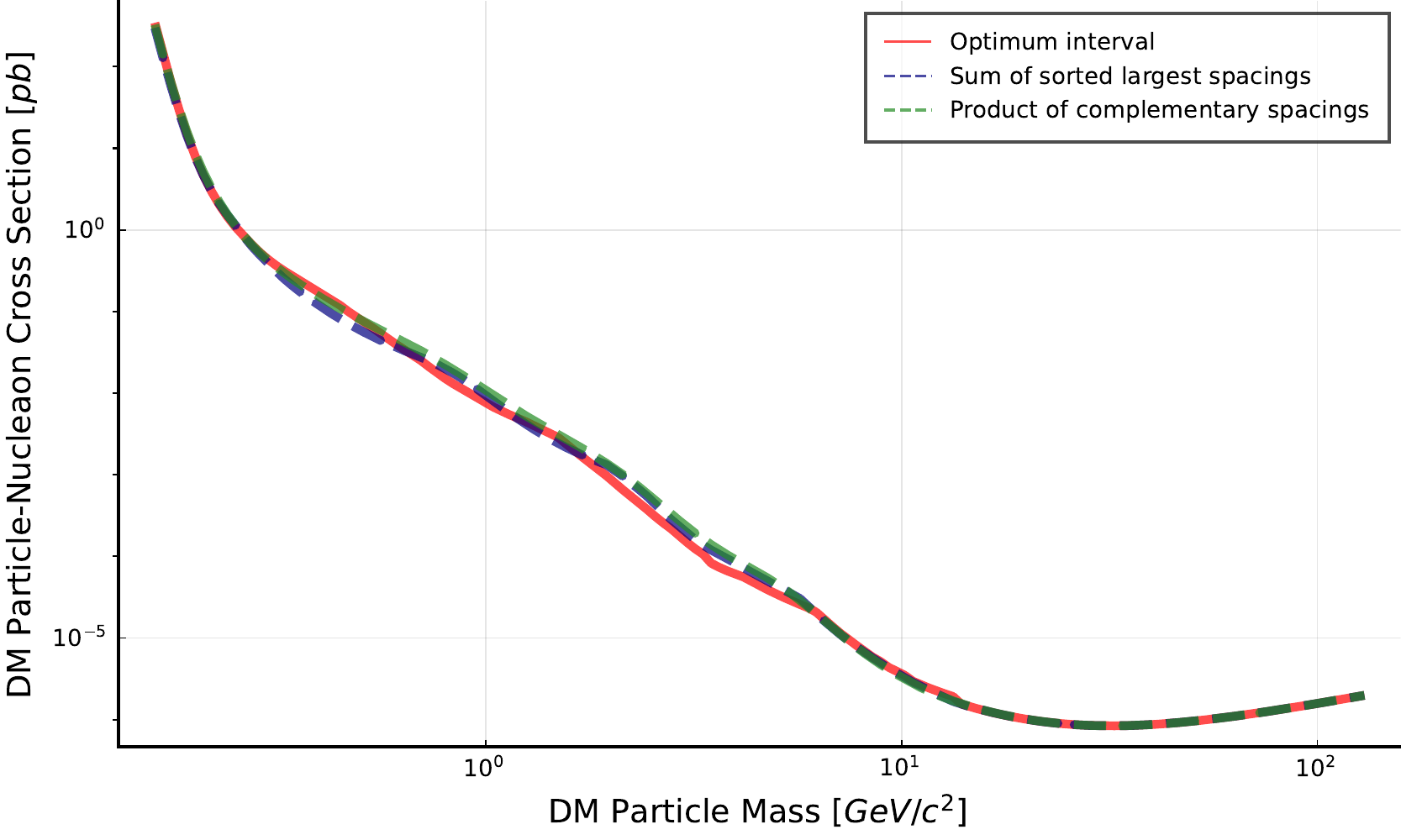}
\caption{$CL=0.90$ upper limit on the WIMP-nucleon cross-section as a function of the WIMP mass calculated with different tests.}
\label{fig:CRESST_limit}
\end{figure}

\begin{figure}[h]
\centering
\includegraphics[width=0.45\textwidth]{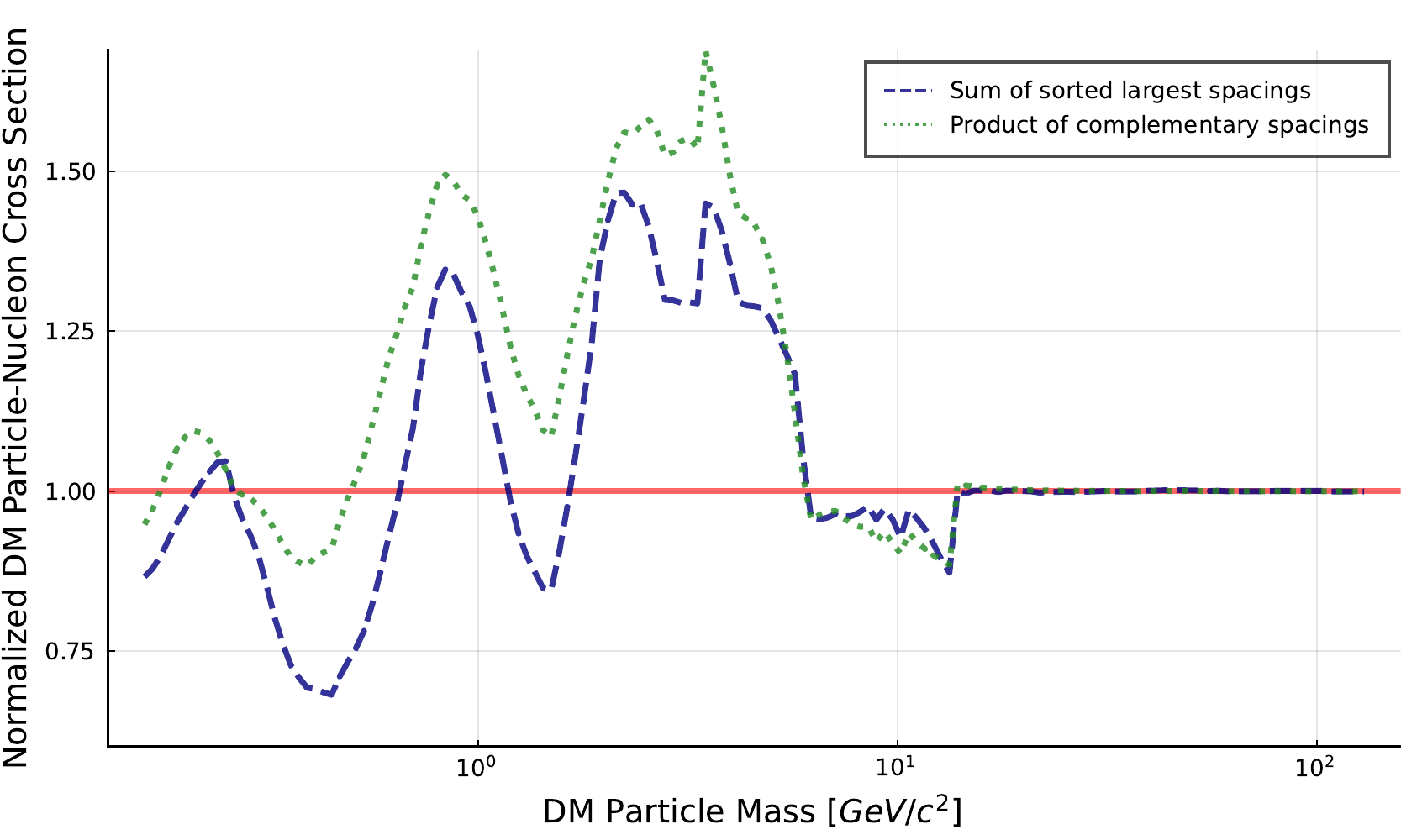}
\caption{$CL=0.90$ upper limit on the WIMP-nucleon cross-section normalized to the Optimum Interval's result as a function of the WIMP mass calculated with our proposed methods.}
\label{fig:CRESST_limit_gain}
\end{figure}

Given a specific WIMP mass and assuming a WIMP velocity distribution it is possible to calculate the differential rate $\frac{dN}{dE}$ across an energy range of interest $[E_{min}, E_{max}]$. Denoting the integral of the differential rate over the whole energy range as $\Lambda = \int_{E_{min}}^{E_{max}} \frac{dN}{dE} dE$, for a given set of ordered events $\{E_i\}$, the probability integral transformation is simply:

\begin{equation}
    x_i = \frac{1}{\Lambda} \cdot \int_{E_{min}}^{E_i} \frac{dN}{dE} dE
\end{equation}

\noindent yielding a set of ordered events $\{x_i\}$ in the unit interval $[0,1]$.

Fig.~\ref{fig:CRESST_data_keV} (bottom) shows the distribution of events in the cumulative space after transforming using the signal distributions calculated at two different WIMP masses.
Apart from small differences, the two datasets are very similar, showing an extremely peaked distribution close to 1 and an almost linear distribution of events in the rest of the unit interval.

Fig.~\ref{fig:CRESST_limit} shows the $CL=0.90$ upper limits on the cross-section we calculated using our methods as well as those computed with the Optimum Interval method, officially used by the CRESST collaboration, which match the officially published limits.

Comparing the results of our calculations we notice there are no large deviations from one another.
To better grasp the differences across results, we normalize the limits we obtain with our methods to the official ones (obtained with the Optimum Interval method), as shown in Fig.~\ref{fig:CRESST_limit_gain}.
Here, we notice that for the given data the Product-of-Complementary-Spacings method yields 25\% to 50\% higher limits on average, whereas the Sum-of-Sorted-Spacings presents an oscillating behavior, being able to provide up to 30\% lower limits for low WIMP masses and up to 40\% higher limits for masses of the order of $\sim 5 \mathrm{Gev/c^2}$.

Finally, for WIMP masses $\geq 20 \mathrm{Gev/c^2}$, all methods saturate and yield the same result, reconstructing a signal event rate of $\sim 2.3$ events, corresponding to the $CL=0.90$ limit of the Poisson test for an empty analysis window. 

This example based on a published data set, as well as the results of our performance comparisons, shows that in general there is no ``best'' test statistic when it comes to setting upper limits in the presence of unknown backgrounds, but the results are highly dependent on the actual distribution of events.

\section*{Conclusions}

\noindent In this paper we present two new methods to set upper limits in the presence of unknown backgrounds, namely the Sum-of-Sorted-Spacings and the Product-of-Complementary-Spacings, and discuss their performance against the Optimum Interval method. The proposed tests leverage the presence of regions in the analysis window with low event density, regardless of their number or location relative to one another in order to estimate the underlying uniform event distribution in the cumulative space. These features allow our tests to be viable alternatives for the analysis of rare process searches that aim to set competitive limits on their parameters of interest, especially when faced with peaked multimodal backgrounds.

\section*{Acknowledgements}

\noindent We thank Dr. Heerak Banerjee, Dr. Oliver Schulz and Dr. Nahuel Ferreiro Iachellini for the helpful discussions and comments that led to finalising this work.

\section*{Appendix A}

\noindent Given $m+2$ ordered samples $\{x_{(i)}\}$ in the interval $[0, \mu]$, where $x_{(0)}=0$ and $x_{(m+1)} = \mu$, we can define the $m+1$ spacings $\{s_{j}\}$ as $s_j = x_{(j)} - x_{(j-1)}$ and the ordered set of spacings $\{s_{(j)}\}$ such that $s_{(j)} < s_{(k)} \forall j<k$ .

The distribution of $s_{(1)}$ is known:

\begin{equation}
    P(s_{(1)} = x|m, \mu) = \frac{m(m+1)}{\mu} \left[ 
1 - \frac{x(m+1)}{\mu} \right]^{m-1}
\label{eq:pdf_min_spacing}
\end{equation}

\noindent The joint distribution of $s_{(1)}$ and $s_{(2)}$ can be written as follows:

\begin{multline}
    P(s_{(1)} = x, s_{(2)} = y|m, \mu) = \\ = P(s_{(1)} = x|m, \mu) \cdot P(s_{(2)} = y|m, \mu, s_{(1)} = x)
\end{multline}

\noindent Given the ordered set $\{s_{(j)}\}$, one can subtract $s_{(1)}$ from all spacings and reduce the set to only $m$ elements $s_{(2)} - s_{(1)},...,s_{(m+1)} - s_{(1)}$ which sum up to $\mu - (m+1) \cdot s_{(1)}$. 
Given this reduces set, $s_{(2)} - s_{(1)}$ is the new smallest elements, whose distributions is given by Eq.~\ref{eq:pdf_min_spacing}.
Thus we can rewrite $P(s_{(1)} = x, s_{(2)}=y)$ as follows:

\begin{multline}
    P(s_{(1)} = x, s_{(2)} = y|m, \mu) = \\ = P(s_{(1)} = x|m, \mu) \cdot P(s_{(1)} = y - x|m-1, \mu - (m+1)x)
\end{multline}

\noindent Given $P(s_{(1)} = x, s_{(2)}=y)$, we can integrate over its support to marginalize $P(s_{(1)} + s_{(2)} = z)$.

We define the sum of the $k$ smallest spacings as $S_k$: $S_k = \sum_{j=1}^k s_{(j)}$.
The joint distribution of $P(s_{(1)}=x, g_k = y)$ can be expressed as:

\begin{multline}
    P(s_{(1)} = x, S_k = y|m, \mu) = \\ = P(s_{(1)} = x|m, \mu) \cdot P(S_k = y|m, \mu, s_{(1)}=x) = \\ = P(s_{(1)} = x|m, \mu) \cdot P(S_{k-1} = y - kx|m-1, \mu - (m+1)x)
\end{multline}

\noindent Assuming we know $P(s_{k-1} = x)$ we can marginalize over $s_{(1)}$ in order to obtain the distribution of $g_k$.
This shows a recursion between the distributions of $S_{k-1}$ and $S_k$, which allows us to formulate an hypothesis on the distribution of $S_k$ which can be proven by induction.
The distribution of $S_k$ we find is:

\begin{multline}
    P(S_k=x|n, 1) = \\ = A(k, n) \cdot \sum_{j=1}^{k} a(j, k) \left[ 1 - \left( \frac{n+2-j}{k+1-j} \right) x \right]^{n-1} h(x,j,k,n)
\end{multline}

\noindent with:

\begin{equation} \label{coeff_A_sumk}
    A(k, n) = \frac{n (n+1)!}{(n+1-k)^{k-1}(n+1-k)!} 
\end{equation}

\begin{equation} \label{coeff_a_sumk}
    a(j, k) = \frac{(-1)^{j-1}(k+1-j)^{k-2}}{(k-j)!(j-1)!}
\end{equation}

\begin{equation} \label{coeff_h_sumk}
    h(x,j, k, n) = H(x) - H \left( x - \frac{k+1-j}{n+2-j} \right)
\end{equation}

\noindent where $H(x)$ is the Heaviside step function.

Finally, since all spacings $s_{(j)}$ sum up to one in the unit interval, the sum of the $m+1-k$ smallest spacings is the complement of the sum of $k$ largest spacings $G_k = \sum_{j=m+2-k}^{m+1} s_{(j)}$.
Thus the distribuiton of $g_k$ is simply:

\begin{equation} \label{p_G_sumk}
    P(G_k = x|m, 1) = P(S_{m+1-k} = 1-x|m, 1)
\end{equation}

\section*{Appendix B}

Using LeCam's theorem \citep{LeCam:1958}, we are able to derive the asymptotic distribution of the PCS test-sstatistic.
The asymptotic distribution of PCS as the number of samples $n \rightarrow \infty$ is:

\begin{equation}
    f_C(C | n \rightarrow \infty) = \mathcal{N}(n \cdot \mu_{\infty},\, n \cdot \sigma_{\infty})
\end{equation}

\noindent where the parameters are given by:

\begin{equation}
    \mu_{\infty}(x) = e^{-x} \left[ \mathrm{E}_1(-x) - 2i \pi \right]
\end{equation}

\begin{multline}
    \sigma^2_{\infty}(x) = e^{-x} \left[ 2A(x) - 4i \pi B(x) -2x e^{-x} \mathrm{E}_1(-x) \right] - \\ - 1- (x^2 + 1) e^{-2x} C(x)
\end{multline}

\noindent where:

\begin{multline}
    A(x) = x F \left( \frac{[1,1,1]}{[2,2,2]}; x \right) + \\ + \frac{\ln^2(-x)}{2} + \gamma \ln(-x) + \frac{\pi^2}{12} + \frac{\gamma^2}{2}
\end{multline}

\begin{equation}
    B(x) = \gamma + \ln(x) + x
\end{equation}

\begin{equation}
    C(x) = \left[ e^{-x} \mathrm{E}_1(-x) \right]^2 + 4i \pi e^{-x} \mathrm{E}_1(-x) - 4 \pi
\end{equation}

\noindent with $F$ being the hypergeometric function and $\gamma$ the Euler–Mascheroni constant.


\bibliography{main}  
\bibliographystyle{PhysRevStyle.bst}

\end{document}